\documentclass[conference,10pt]{IEEEtran}
\IEEEoverridecommandlockouts
\usepackage{amsmath,amssymb,amsfonts}
\usepackage{algorithmic}
\usepackage{graphicx}
\usepackage{textcomp}
\usepackage{xcolor}
\usepackage{comment}
\usepackage[numbers]{natbib}

\usepackage{hyperref}
\usepackage{mathtools}
\usepackage{multirow}
\usepackage{array}
\usepackage{colortbl}
\usepackage{footnote}
\usepackage{tablefootnote}
\definecolor{mygray}{gray}{.9}
\usepackage{pifont}

\usepackage{lineno}

\def\BibTeX{{\rm B\kern-.05em{\sc i\kern-.025em b}\kern-.08em
    T\kern-.1667em\lower.7ex\hbox{E}\kern-.125emX}}
\begin{document}
%\include{latex/macros}
%Including images in your LaTeX document requires adding
%additional package(s)

\title{\huge{Ensembler: Protect Collaborative Inference Privacy from Model Inversion Attack via Selective Ensemble}
%Ensembler: Ensemble-Based Model Confusion as Defense during Private Inference
%Ensembler: Combating Model Inversion Attacks Using Model Ensemble During Collaborative Inference
%\\
%{\footnotesize \textsuperscript{*}Note: Sub-titles are not captured for https://ieeexplore.ieee.org  and
}
%	Dancheng Liu, Chenhui Xu, Jiajie Li, Amir Nassereldine and Jinjun Xiong

\author{\IEEEauthorblockN{Dancheng Liu}
\IEEEauthorblockA{\textit{Computer Science and Engineering} \\
\textit{University at Buffalo}\\
Buffalo, USA \\
dliu37@buffalo.edu}
\and
\IEEEauthorblockN{Chenhui Xu}
\IEEEauthorblockA{\textit{Computer Science and Engineering} \\
\textit{University at Buffalo}\\
Buffalo, USA \\
cxu26@buffalo.edu}
\and
\IEEEauthorblockN{Jiajie Li}
\IEEEauthorblockA{\textit{Computer Science and Engineering} \\
\textit{University at Buffalo}\\
Buffalo, USA \\
jli433@buffalo.edu}
\and
\IEEEauthorblockN{Amir Nassereldine}
\IEEEauthorblockA{\textit{Computer Science and Engineering} \\
\textit{University at Buffalo}\\
Buffalo, USA \\
amirnass@buffalo.edu}
\and
\IEEEauthorblockN{Jinjun Xiong}
\IEEEauthorblockA{\textit{Computer Science and Engineering} \\
\textit{University at Buffalo}\\
Buffalo, USA \\
jinjun@buffalo.edu}
}
\maketitle

\begin{abstract}
%As edge-based automatic speech recognition (ASR) technologies become increasingly prevalent for the development of intelligent and personalized assistants, three important challenges must be addressed for these resource-constrained ASR models, i.e., adaptivity, incrementality, and inclusivity. We propose a novel ASR framework, PI-Whisper, in this work and show how it can improve an ASR’s recognition capabilities adaptively by identifying different speakers’ characteristics in real-time, how such an adaption can be performed incrementally without repetitive retraining, and how it can improve the equity and fairness for diverse speaker groups. More impressively, our proposed PI-Whisper framework attains all of these nice properties while still achieving state-of-the-art accuracy with up to 13.7\% reduction of the word error rate (WER) with linear scalability with respect to computing resources.

%Deep learning models have exhibited remarkable performance across various domains, but the burgeoning model sizes compel edge devices to offload a significant portion of the inference process to the cloud. While this practice offers numerous advantages, it also raises critical concerns regarding user data privacy. In scenarios where the cloud server's trustworthiness is in question, the need for a practical and adaptable method to safeguard data privacy becomes imperative.

For collaborative inference through a cloud computing platform, it is sometimes essential for the client to shield its sensitive information from the cloud provider. In this paper, we introduce Ensembler, an extensible framework designed to substantially increase the difficulty of conducting model inversion attacks by adversarial parties. Ensembler leverages selective model ensemble on the adversarial server to obfuscate the reconstruction of the client's private information. Our experiments demonstrate that Ensembler can effectively shield input images from reconstruction attacks, even when the client only retains one layer of the network locally. Ensembler significantly outperforms baseline methods by up to 43.5\% in structural similarity while only incurring 4.8\% time overhead during inference.
\end{abstract}

%\begin{IEEEkeywords}
%Privacy, Deep learning, Ensemble learning
%\end{IEEEkeywords}

\section{Introduction}

\begin{figure}[t]
    \centering
    \includegraphics[width=\linewidth]{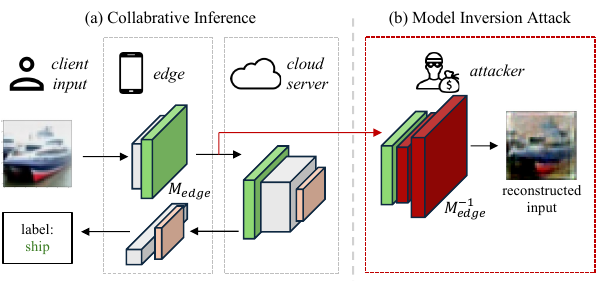}
    \caption{An illustration of (a) collaborative inference, where the client offloads computation to the server; and (b) model inversion attack, where the adversarial server builds a decoder $M_{edge}^{-1}$ to decode the client's input from the intermediate features computed by the client's private network $M_{edge}$. }
    \label{fig:shredder1}

\end{figure}

Deep learning models have achieved exceptional performance across critical domains \cite{IMAGENET}, but their increasing size poses challenges for computationally constrained edge devices like mobile phones \cite{complexity2021, qin2024empirical}. 
%Therefore, reducing the computational workload on these resource-limited edge devices becomes imperative.
When the model's size and complexity exceed the computational ability of the edge devices, a prevalent solution to reduce the workload on these resource-limited edge devices is through collaborative inference (CI) \cite{He2019}, as shown in the left side of Fig. \ref{fig:shredder1}. During inference, the client (edge device) computes the first few layers of the network and sends the intermediate features to the server (cloud), where the server computes the computationally expensive layers and returns the features or the results back to the client. However, this raises concerns about the privacy of sensitive client data, especially when the cloud server may be adversarial and tasks involve sensitive information like medical or facial data \cite{preva}. Consequently, there is a growing need for secure, accurate, and efficient machine learning frameworks that enable privacy-preserving collaboration between edge devices and cloud servers.

During CI, a classical yet powerful attack on the client's sensitive data is through model inversion attacks (MIA). As shown in the right side of Fig. \ref{fig:shredder1}, an adversarial server aims to inverse the client's private network to reconstruct the private input through intermediate features. We will elaborate on the details of MIA in Section \ref{background}. He et. al. show that the client's data is especially vulnerable when the client retains only one single convolutional layer \cite{He2019}.  While the client can potentially keep more privacy as it keeps more layers, such a method is less practical in the real world due to the limitations of the computational power of edge devices. Shredder \cite{He2019} uses a noise injection layer before the client sends out computed results to reduce mutual information between client and server while maintaining good classification accuracy. Nevertheless, Preva \cite{preva} demonstrates that Shredder falls short in safeguarding facial images from recovery. Through experimentation with the noise injection layer proposed by Shredder on a shallow layer, we observed significant accuracy drops with combined multiplicative and additive noise. On the other hand, simple additive noise resulting in approximately a 3\% drop in accuracy failed to protect images from recovery.%, as shown in the reconstructed image in Fig. \ref{fig:shredder1}.
%Lu proposed to use a policy-based processor between client and server to protect private information, but figures in their work seem to indicate that the effectiveness of their policy should be attributed to removing some regions from the original image that contain sensitive data. While such an approach is effective in some cases, it falls short in scenarios where sensitive information is embedded within the image, such as in facial authentication tasks.

%, applied to a ResNet \cite{he2016deep} architecture on CIFAR-10 \cite{krizhevsky2009learning},
\begin{comment}

\begin{figure}{}
    \centering
    \includegraphics[scale=0.4]{./img/original.png}
    \includegraphics[scale=0.4]{./img/recovered.png}
    \caption{Sample image from CIFAR-10 under the protection of Shredder. Left is the original image and right is the recovered image.}
    \label{fig:shredder1}

\end{figure}
\end{comment}

The failure of Shredder \cite{shredder} could be attributed to two factors: the client's computational limit and the server's knowledge of the partial weights. To ensure the best defense, Shredder requires the noise injection layer to be very deep, but that results in the client taking heavy computation work, which is infeasible in real-world scenarios. Meanwhile, a shallow noise injection layer leaves most weights to the server, and with the rich information available in the public model weights, the server can learn a strong and accurate representation of the private network on the client, which leads to a successful MIA. Thus, the key challenge behind a practical CI framework is to defend against an adversarial server's reconstruction ability when the client only holds a minimal portion of the network.

To address the current deficiencies in privacy-preserving edge-cloud collaborative inference scenarios, in this work, we address the aforementioned challenge through the mechanism of ``selective ensemble''. When the reconstruction ability of the adversarial server is paramount, it would be wiser to rather hide the client's private model than defend it directly against the model reconstruction attack. We propose to leverage model ensemble on the cloud and secret private masking on the user side to achieve a ``Schrödinger's model'', where an arbitrary reconstruction attack will lead to a successful, yet different, shadow network that fails to emulate the client's network. 

In this paper, synthesizing the above analysis, we propose \textit{Ensembler}, a novel secure CI framework designed to substantially increase the effort required to recover client input.
The proposed architecture consists of multiple parallel networks, but only a subset of them is actually activated by the client during inference. An arbitrary MIA on the network will lead to a seemingly successful reconstruction, but such reconstruction does not represent the real client's weights.

\noindent\textbf{Contributions} We make the following contributions:
\begin{itemize}
    \item We identify and address the key challenge in practical private collaborative inference: when the model split location between the client and the server is shallow, a direct defense fails to work.
    \item We reveal that a secret selection of a subset of the networks in an ensemble architecture can be useful for defending against model inversion attacks.
    \item We propose \textit{Ensembler}, which is not only a standalone framework that significantly increases the adversary server's reconstruction difficulty but can also be seamlessly integrated with existing complex algorithms to construct practical and secure inference architectures.
\end{itemize}

To verify the effectiveness of \textit{Ensembler}, we conduct experiments on multiple datasets.
%we conduct experiments using ResNet-18 \cite{he2016deep} on three datasets: CIFAR-10 \cite{krizhevsky2009learning}, CIFAR-100 \cite{krizhevsky2009learning}, and a subset of Celeb-HQ \cite{zhu2022celebvhq}. 
Compared to models without using model ensemble, when defending against MIA, \textit{Ensembler} achieves up to 43.5\% decrease in structural similarity and 40.5\% decrease in peak signal noise ratio. In addition, compared to the previous popular approach, Shredder \cite{shredder}, \textit{Ensembler} better protects the client's sensitive information with a 10.7\% decrease in peak signal noise ratio, highlighting its effectiveness even as a standalone framework.

\section{Background}
\label{background}
\subsection{Collaborative Machine Learning}
The development of mobile graphic processing units (GPUs) has ushered in a new era where machine learning tasks are increasingly deployed with a portion of the computation being handled by edge devices. Related areas include federated learning, where multiple edge devices jointly train a DL model \cite{Async-FHL,Yaldiz2023}; split learning, where a DL model is split into two or more parts, and the client and server jointly train it \cite{SL2019}; and collaborative inference, where a DL model is split, with only a portion deployed on the server to provide services \cite{He2019, DPFE}.
In this paper, we will focus on the inference part and assume that the training phase of DL models is secure. Though the training phase is sometimes also susceptible to adversarial attacks aimed at stealing sensitive information \cite{Li_2022_CVPR, Zhang2021train},  private inference is still more prevalent in most practical scenarios.

There have been multiple works addressing this formidable challenge of safeguarding the client's sensitive information in collaborative inference scenarios. \cite{survey_xu,pmlr-v235-xu24ae} provides an extensive discussion on different algorithmic and architectural choices and their impacts on privacy protection. For simplicity and without loss of generalization, we will group existing approaches into two categories in this paper: encryption-based algorithms that guarantee privacy at the cost of thousands of times of time efficiency \cite{Delphi,rathee2020cryptflow2,lam2024efficient,li2024xmlp}, and perturbation-based algorithms that operate on the intermediate layers of a DL architecture, introducing noise to thwart the adversary's ability to recover client input \cite{shredder,DPFE,preva, warit2020gan}. Since perturbation-based algorithms directly operate on the intermediate outputs from the client, they incur minimal additional complexity during the inference process. As opposed to guaranteed privacy provided by encryption-based algorithms, perturbation-based algorithms suffer from the possibility of privacy leakage, meaning sensitive private information may still be recoverable by the adversarial server despite the introduced perturbations, but they are nonetheless more practical to use due to the less additional costs associated with the methods.

\subsection{Threat model}
\label{threat_model}
In this paper, we consider the collaborative inference task between the client and the server, which acts as a semi-honest adversarial attacker that aims to steal the raw input from the client. Formally, we define the system as a collaborative inference on a pre-trained DNN model, $\mathbf{M}(x,\theta)$, where the client holds the first and the last few layers (i.e. the ``head" and ``tail" of a neural network), denoted as $\mathbf{M}_{c,h} (x,\theta_{c,h})$ and $\mathbf{M}_{c,t} (x,\theta_{c,t})$. The rest of the layers of DNN are deployed on the server, denoted as $\mathbf{M}_{s} (x,\theta_{s})$. $\theta$ is the trained weights of $M$, where $\theta=\{\theta_{c,h},\theta_{s},\theta_{c,t}\}$. The complete collaborative pipeline is thus to make a prediction of incoming image $x$ with $\mathbf{M}_{c,t}[\mathbf{M}_{s}[\mathbf{M}_{c,h}(x)]]$.

During the process, the server has access to $\theta_{s}$ and the intermediate output $\mathbf{M}_{c,h}(x)$. In addition, we assume that it has a good estimate of the DNN used for inference. That is, it has auxiliary information on the architecture of the entire DNN, as well as a dataset in the same distribution as the private training dataset used to train the DNN. However, it does not necessarily know the hyper-parameters, as well as the engineering tricks used to train the model. Since the server is a computation-as-a-service (CaaS) provider, it is assumed to have reasonably large computation resources. While it is powerful in computing, the server is restricted from querying the client to receive a direct relationship between raw input $x$ and intermediate output $\mathbf{M}_{c,h}(x)$.

To reconstruct raw input $x$ from the intermediate output $\mathbf{M}_{c,h}(x)$, the server adopts a common model inversion attack \cite{He2019,preva,Dosovitskiy2016autoencoder}. To obtain $x$'s reconstruction, MIA aims to obtain the inverse of $\mathbf{M}_{c,h}$, which will be denoted as $\mathbf{M}_{c,h}^{-1}$. The server constructs a shadow network $\tilde{\mathbf{M}}(x,\tilde{\theta_{c,h}}, \theta_{s}, \tilde{\theta_{c,t}}) :\{\tilde{\mathbf{M_{c,h}}}, \mathbf{M}_{s}, \tilde{\mathbf{M_{c,t}}}\}$ such that $\tilde{\mathbf{M}}$ simulates the behavior of $\mathbf{M}$. After training $\tilde{\mathbf{M}}$, the adversarial server is able to obtain a representation $\tilde{\mathbf{M_{c,h}}}$ such that $\tilde{\mathbf{M_{c,h}}}(x) \sim \mathbf{M}_{c,h}(x)$. As the next step, with the help of a decoder, $\tilde{\mathbf{M_{c,h}^{-1}}} \sim \mathbf{M}_{c,h}^{-1}$, to reconstruct the raw image from intermediate representation, it is able to reconstruct the raw input from $\mathbf{M}_{c,h}(x)$.

\subsection{Assumptions of other related works}
In this section, we provide an overview of various attack models and the assumptions adopted in other works related to collaborative inference (CI) under privacy-preserving machine learning (PPML). Since different works potentially use different collaboration strategies between the client and the server, we will use the generic notation, where $\mathbf{M}_c$ is held by the client, and $\mathbf{M}_s$ is held by the server. Generally, the attacks from the server will fall into three categories: 

% Typically, works in this area consider a system where a trained neural network, denoted as $\mathbf{M}(x,\theta)$, is associated with a particular party, often the client ($C$).  At the same time, there exists an adversarial party, which is often a (compromised) server ($S$) that wants to learn certain sensitive or private information related to $\mathbf{M}$. Collaborative inference occurs when $C$ tasks $S$ with a portion of the inference pipeline. Under CI, $\mathbf{M}(x,\theta)$ is split between C and S, denoted as $\mathbf{M}_c$ and $\mathbf{M}_s$, such that $\mathbf{M}(x)=\mathbf{M}_s[\mathbf{M}_c(x)]$. 

\begin{itemize}
    \item \textbf{Membership Inference Attacks} that try to predict if certain attributes, including but not limited to individual samples, distributions, or certain properties, are a member of the private training set used to train $\mathbf{M} (x, \theta)$. If successful, the privacy of the training dataset will be compromised. We refer readers to the survey by \cite{membership_survey} and \cite{Salem2023dataset} for more details.
    \item \textbf{Model Inversion Attacks} that try to recover a particular input during inference when its raw form is not shared by the client. For example, in an image classification task, the client may want to split $\mathbf{M}$ such that it only shares latent features computed locally to the server. However, upon successful model inversion attacks, the server will be able to generate the raw image for classification tasks based on the latent features. It is important to note that, in this paper,  we adopt the same definition of model inversion attacks as of \cite{He2019}. This term also refers to attacks that reconstruct the private training dataset in other works. We will focus on reconstructing private raw input for the rest of this paper.
    \item \textbf{Model Extraction Attacks} that try to steal the parameters and even hyper-parameters of $\mathbf{M}$. This type of attack compromises the intellectual property of the private model and is often employed as sub-routines for model inversion attacks when the server lacks direct access to the private model's parameters.
\end{itemize}

Different works also make different assumptions about the capability of the server. First, it is widely accepted that the server has sufficiently yet reasonably large computing power and resources, as its role is often providing ML service. Regarding the auxiliary information on $\mathbf{M}$, they generally fall into three levels:
\begin{itemize}
    \iffalse
    \item \textbf{White Box} assumes that $S$ has full access of $\mathbf{M}$. It has complete information on the structure, and parameters of $\mathbf{M}$, as well as the training dataset used to train $\mathbf{M}$ \cite{Liu2021Survey}. This setting is often associated with attacks that try to reconstruct private training datasets \cite{Wang2022Ensemble_inversion,Zhang2020Secret_Revealer,Haim2022Invert_data}. \JX{This seems to be contradictory - it had just said that we assumed that the attacker knows the training dataset.}
    \fi
    %%% new white box definition
    \item \textbf{White Box} assumes that the server has full access to architecture details of $\mathbf{M}$ such as the structure and parameters \cite{Liu2021Survey}. Different definitions also add different auxiliary available information such as training dataset \cite{Liu2021Survey}, corrupted raw input \cite{Zhang2020Secret_Revealer}, or a different dataset \cite{Wang2022Ensemble_inversion}. This setting is often associated with attacks that try to reconstruct private training datasets \cite{Wang2022Ensemble_inversion,Zhang2020Secret_Revealer,Haim2022Invert_data}.    
    \item \textbf{Black Box} assumes that the server does not have any information on both $\mathbf{M}$ and the training dataset. However, it is allowed to send unlimited queries to the client to get $\mathbf{M}_c(x)$ \cite{Xu20203black,Kahla2022black_label}.
    \item \textbf{Query-Free} restricts the server from querying $\mathbf{M}_c$. While such an assumption greatly limits the reconstruction ability of the adversarial party, there are no limitations on auxiliary information available to the server besides the actual weights of $\mathbf{M}_c$. \cite{He2019, ding2023patrol} have both shown that $\mathbf{M}_c$ is still vulnerable to leaking private information of its input when the server has information on the model architecture and training dataset. Our work uses this setting.
\end{itemize}

\begin{figure*}[t]
    \centering
    \includegraphics[trim=0 0 0 0, clip, width=\linewidth]{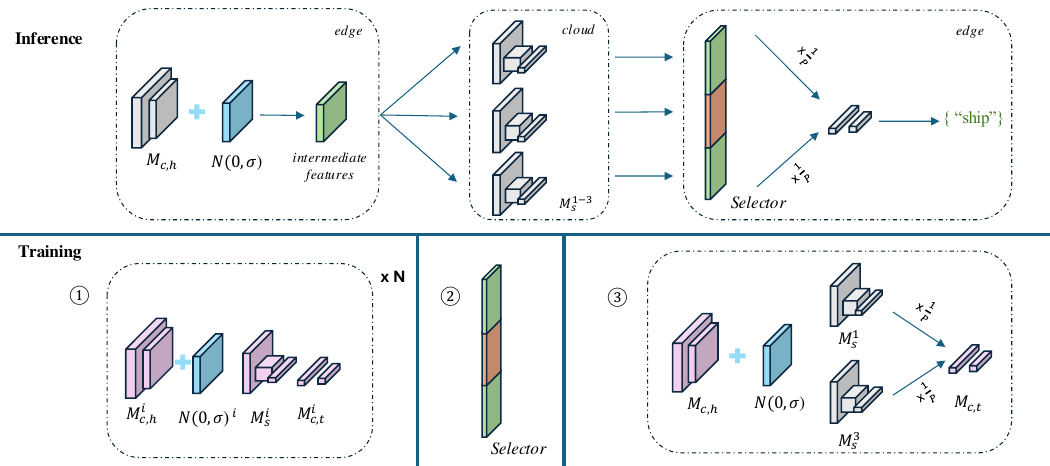}
    \caption{Illustration of the proposed architecture, \textit{Ensembler}, during inference and training. Unlike traditional CI pipelines, during inference, \textit{Ensembler} deploys $N$ neural networks on the server and uses a private selector to activate $P$ of the $N$ nets, making the reconstruction of the client network infeasible. During training, \textit{Ensembler} follows a three-stage training procedure, and the purple elements are trainable parameters of each stage.}
    \label{architecture}
\vspace{-3mm}
\end{figure*}
\section{\textit{Ensembler} architecture}
\label{arch}

While it is possible to protect sensitive data by increasing the depth of the client network, such depth is often impractical for edge devices due to the computational demands involved. In this section, we present \textit{Ensembler}, a framework that augments the privacy of intermediate information sent by the client without requiring extra computation efforts of the client during inference. \textit{Ensembler} is highly extensible, and it is compatible with existing works that apply noises and perturbation during both DNN training and inference.  We will go over the inference pipeline of \textit{Ensembler} in Section \ref{overview}, the training stage of this new framework in Section \ref{training}, an intuitive explanation of why \textit{Ensembler} works in Section \ref{intuition}, and finally the time complexity analysis in Section \ref{time}.

\subsection{Inference Pipeline}
\label{overview}
As illustrated in Figure \ref{architecture}, \textit{Ensembler} leverages model ensemble on the server to generate a regularized secret $\mathbf{M}_{c,h}$ that is hard to be reconstructed by the adversarial server. It comprises three parts: standard client layers, $N$ different server nets,
and a Selector. During the CI pipeline, the client computes the addition of $\mathbf{M}_{c,h}(x)$ and a predefined noise, and transmits the intermediate output to the server. The server then feeds the intermediate output through each of the $\mathbf{M}_{s}^i$ and reports the output of each $\mathbf{M}_{s}^i$ to the client. The client then employs a user-defined private selector to perform a selection of the feedback from the server, which activates the results of $P$ out of $N$ nets and combines them. As a final step, it computes the last few layers ($\mathbf{M}_{c,t}(x)$) to classify the input image. We will introduce these separately in this section.

\subsubsection{Client Layers}
During collaborative inference, the client runs a part of the DNN. Under the proposed framework, the client is responsible for running the first $h$ layers $\mathbf{M}_{c,h}$ and the last $t$ layers $\mathbf{M}_{c,t}$. These layers are the same as the client part of a typical collaborative inference framework. $\mathbf{M}_{c,h}$ takes the raw input (often an image) and outputs the intermediate result, whereas $\mathbf{M}_{c,t}$ takes the output from the server as input and outputs the likelihood of each class.

\subsubsection{Server Nets}
On the server side, the network consists of $N$ copies of DNN, with each $M_{s}^i \in M_{s}$ corresponding to what the server would normally process in a typical CI pipeline. That is, each $M^i: \{M_{c,h}, M_{s}^i, M_{c,t}\}$ is a standard pipeline for the inference task. Upon receiving $\mathbf{M}_{c,h}(x)+N(0,\sigma)$, which is the intermediate features generated by the client, the server shall feed this input into each of $M_{s}^i$. It outputs $N$ representations of hidden features used for classification and sends all of them back to the client. We denote those outputted feature maps as $M_{s}(x') = \{M_{s}^i(x')\}$.

\subsubsection{Selector and $\mathbf{M}_{c,t}$}
To increase the difficulty of the server reconstructing the model and recovering the raw input, a user-defined private selector is applied to the outputs returned by the server before the client runs $\mathbf{M}_{c,t}$. The selector is a secret activation function, activating a subset of $M_{s}(x')$. We denote this subset of activated feature maps as $M_{s}(x')_p$, and there are in total $P$ activated feature maps. Then, the Selector is defined by Eq. \ref{select}, where $S_i$ is the activation from the Selector, and $\odot$ is the element-wise multiplication. For simplicity, we consider $S_i=1/P$. In other words, the client will normalize $P$ of the $N$ feature maps returned by the adversarial server and concatenate them as the input features to $\mathbf{M}_{c,t}$ while keeping those selections private to the server.

\begin{equation}
    \label{select}
    Sel[M_{s}(x)] = Concat [S_i \odot f \quad \forall f \in M_{s}(x')_p ]
\end{equation}

\subsection{Intuition behind \textit{Ensembler}}
\label{intuition}
In this section, we provide an intuitive explanation of \textit{why} the proposed architecture can defend the client's private information against MIA through a simulated MIA performed by the adversarial server. As defined by the threat model in Section \ref{threat_model}, the server has access to the training dataset, model architecture, and $M_{s}$. To perform a MIA, it will construct $\tilde{\mathbf{M_{c,h}}}$ and $\tilde{\mathbf{M_{c,t}}}$. Now, let's further assume that each of $M_{s} \in \{M_{s}^i\}$ has different weights (how to achieve this will be shown in the next subsection). 
% Without brute force trying all combinations, and since it does not know the user-defined selector, the adversarial server remains two possible MIA constructions.
With brute-forcing all combinations being exponential and unpractical, and since the adversarial server does not know the user-defined selector,  there remain two possible MIA constructions.
The server could either construct the shadow network with a random subset of $\{M_{s}^i\}$, or it could use the entire $\{M_{s}^i\}$. We will show that both methods result in $\tilde{\mathbf{M_{c,h}}}(x) \not\sim \mathbf{M}_{c,h}(x)$. Without loss of generality, for the first approach, the server is assumed to construct $\tilde{\mathbf{M_{c,h}}}$ from a single $M_{s}^i$ that is a subset of the networks selected by the Selector.

\noindent\textbf{Proposition 1:} Reconstructing input from single neural network $\mathbf{M}_{s}^i$ results in $\tilde{\mathbf{M}_{c,h}}(x) \not\sim \mathbf{M}_{c,h}(x)$. 

Let's consider the most extreme case, where the Selector selects $\mathbf{M}_{s}^1$ and $\mathbf{M}_{s}^2$, and the adversarial server is constructing its shadow network with $\mathbf{M}_{s}^1$. Since $\mathbf{M}_{c,h}$ is trained with both networks, the gradient of weights of $\mathbf{M}_{c,h}$ is calculated as (roughly) $\eta(\nabla_x\mathbf{M}_{s}^1+\nabla_x\mathbf{M}_{s}^2)$ during backpropogation. After minimization, as long as $\nabla_x\mathbf{M}_{s}^2$ is not zero, the weights of $\mathbf{M}_{c,h}$ will be affected by both $\mathbf{M}_{s}^1$ and $\mathbf{M}_{s}^2$ (we will show how to let the gradient not be zero through loss regularization in the next section). Since $\mathbf{M}_{s}^1$ and $\mathbf{M}_{s}^2$ have different weights, their gradient will point to different directions. A shadow network $\tilde{\mathbf{M}_{c,h}}$ trained on only $\mathbf{M}_{s}^1$ will represent different weights compared to $\mathbf{M}_{c,h}$, and thus the decoder based on $\tilde{\mathbf{M}_{c,h}}$ will also reconstruct incorrect client's secret input.
%For any shadow network obtained through single $\mathbf{M}_{s}^i$ , it needs to first simulate the behavior of $\mathbf{M}_{c,h}^i$. In this case, if there exists some $\mathbf{M}_{c,h}^i$ that simulates $\mathbf{M}_{c,h}$, the training loss of the second training phrase is not optimized (Equation. \ref{train2}) due to the regularization term.

\noindent\textbf{Proposition 2:} Reconstructing input from the entire set $\{\mathbf{M}_{s}\}$ results in $\tilde{\mathbf{M}_{c,h}}(x) \not\sim \mathbf{M}_{c,h}(x)$.

A similar conclusion from the analysis in the first case may be derived. Let's still use the simple example of $\mathbf{M}_{s}^1$ and $\mathbf{M}_{s}^2$. This time, the Selector selects only $\mathbf{M}_{s}^1$, but the server decides to train on both $\mathbf{M}_{s}^1$ and $\mathbf{M}_{s}^2$. It is not hard to see that the weights of $\mathbf{M}_{c,h}$ and $\tilde{\mathbf{M}_{c,h}}$ are just swapped with each other, and now, the decoder is still based on incorrect weights and still leads to an incorrect reconstruction of $\mathbf{M}_{c,h}(x)$.

%Since the attacker will construct shadow networks to simulate the behavior of client's private networks, the exact purpose of the two-staged training algorithm is to ensure that the attacker is not able to learn the selector with its shadow network. Through the first stage of training, N different models that have distinctive weights are obtained, yet all of them are able to make comparative predictions on the dataset. An arbitrary ensemble of P out of the N networks will form a new network, whose $M_{c,h}$ will be distinctive from networks under a different combination. That is, since $M_{s}^{i+j}$ would be different from $M_{s}^{i+k}$, $M_{c,h}^{i+j}$ obtained from $M_{s}^{i+j}$ would be different from $M_{c,h}^{i+k}$ obtained from $M_{s}^{i+k}$, where $+$ is ensemble of server nets. Thus, with N networks in the first stage of the algorithm, we will have $2^N$ different possible $M_{c,h}$ that could be the valid answer to the shadow network. When the attacker tries to train an adaptive attacker, the shadow network will learn an arbitrary representation $\tilde{\mathbf{M_{c,h}}}$ and an arbitrary $\tilde{S}$. Such combination is a valid choice in terms of classification accuracy but is nonetheless incorrect compared to the actual $\mathbf{M_{c,h}}$.

\subsection{Training Stages}
\label{training}

As mentioned above, the key idea behind \textit{Ensembler} is to use $N$ \textbf{different} networks with a private Selector to obfuscate MIA. To achieve the $N$ different networks and make sure $M_{c,h}$ is properly trained with all of the selected subsets, the proposed architecture uses a three-staged training pipeline, as shown in the bottom half of Fig. \ref{arch}. 

\noindent\textbf{Stage 1:} First, \textit{Ensembler} obtains $N$ different networks by injecting $N$ different noises after each of the network's $M_{c,h}^i$ and training each of the networks. Since randomly initialized noises are quasi-orthogonal to each other and each network needs to adjust its weights to reverse the effects of the added noise on the intermediate feature map, the resulting weights of $M_{c,h}^i$ will be different from each other. Formally, the first stage optimizes Eq. \ref{train1}, where $N(0,\sigma)^i$ is a fixed Gaussian noise added to the intermediate output. We choose Gaussian noise because of simplicity in implementation, and we'd argue that any method that will lead to distinctive $M_{c,h}$ will be sufficient for this step.
\vspace{-1mm}
\begin{equation}
    \label{train1}
    L_{\theta}^i = -\sum_j y_j * log \text{   } M_{c,t}^i(M_{s}^i[M_{c,h}^i(x)+N(0,\sigma)^i])_j
\end{equation}
\vspace{-1mm}

\noindent\textbf{Stage 2:} After the first training stage, $N$ different DNNs are obtained. The client secretly selects $P$ out of the $N$ networks as the activation of the Selector. 

\noindent\textbf{Stage 3:} With the Selector defined, the client is now able to aggregate $P$ networks together to form the ``ensembled" inference pipeline. As the last stage of the training process, $P$ server networks together form the body of the final network, and their weights are frozen. The client re-trains its private network $M_{c,h}$ and $M_{c,t}$, with a newly defined Gaussian noise added to the output features of $M_{c,h}$. This Gaussian noise follows the same definition as the Gaussian noise in Stage 1's training. Formally, this stage's training loss follows Eq. \ref{train2}. adds a high penalty to the model if the gradient descends only to the direction of some single server net $M_{s}^i$. It is noteworthy that during training this loss has two terms. The first part (in the first line) is the standard cross-entropy loss used to train a classification network, and the second part is a regularization term to enforce $M$ to learn a joint $M_{c,h}$ and $M_{c,t}$ representation from all of the $P$ server networks. In the equation, CS is the cosine similarity, and $\lambda$ is a hyperparameter controlling the regularization strength. The idea behind such a regularization is that if $M_{c,h}$ only learns from one single network, then its weights will be very similar to one of the previous $M_{c,h}^i$. Thus, we regularize it to be as quasi-orthogonal to all of the previous $M_{c,h}^i$ as possible.
\vspace{-3mm}
\begin{equation}
    \label{train2}
    \begin{split}
    &L_{\theta} = \sum_i^{i\in P} CE+ Regularization,\quad where  \\
    &CE = -\sum_j [y_j * log M_{c,t}(Sel[M_{s}^i[M_{c,h}(x)+N(0,\sigma)]])_j] \\
    &Regularization =
    \lambda max[CS(M_{c,h}(x),M_{c,h}^i(x))]
    \end{split}
\end{equation}
\vspace{-5mm}

\subsection{Time Complexity of \textit{Ensembler}}
\label{time}

From the previous subsections, it is clear that the inference time complexity of the proposed framework on the client is the same as the standard non-private inference pipeline. On the server, it is N times the individual network on a single-core GPU, and there is negligible extra communication cost between the client and the server. However, it is noteworthy to emphasize that since each $M_s^i$ is independent of each other, the proposed framework is friendly to parallel execution and even multiparty (multi-server) inference. Under those settings, the theoretical time complexity of $O(N)$ would be replaced with lower practical time costs. 

Regarding the MIA's time complexity, following the analysis in Section \ref{intuition}, it is not hard to see that the expected MIA time of a single server is $O(2^N)$. Since an arbitrary reconstruction on some subset of the network will lead to a reasonable shadow network $\tilde{\mathbf{M_{c,h}}}$, the server has no way of telling whether its reconstruction is an actual representation of the client's network. Thus, to ensure optimal reconstruction, it must use brute force and try all possibilities. In addition, under multiparty inference scenarios, without access to a subset of the activated networks, the adversarial server will not be able to perform the desired reconstruction.

\section{Experiments and Evaluations}
\label{experiment}
\begin{table*}[t]
    \centering
    \tabcolsep 10pt
    \caption{Defense quality of \textit{Ensembler} and Single \cite{dwork2006} across datasets. For SSIM and PSNR, lower values indicate better defense quality. We report on the two reconstruction attacks discussed in Section \ref{intuition}. For reconstructing with a single network, we report the strongest reconstruction attack (least defense quality) out of $N$ networks.}
    %The bolded values are the best reconstruction from the MIA attack (defense in worst-case scenarios).}
    \label{summary_all}
    \begin{tabular}{l c c c c c c c c c}
    \hline
    & \multicolumn{3}{c}{CIFAR-10} & \multicolumn{3}{c}{CIFAR-100} & \multicolumn{3}{c}{CelebA-HQ} \\
    \cline{2-10}
    Name & $\Delta$Acc & SSIM $\downarrow$ & PSNR $\downarrow$ & $\Delta$Acc & SSIM $\downarrow$ & PSNR $\downarrow$ & $\Delta$Acc & SSIM $\downarrow$ & PSNR $\downarrow$ \\ \hline
    Single & 2.15\% & 0.39 & 7.53 & -0.97\% & 0.46 & 8.52 & -1.24\% & 0.27 & 14.31 \\ 
    Ours - Adaptive & -2.13\% & 0.06 & 5.98 & 0.31\% & 0.09 & 4.77 & 2.39\% & 0.09 & 13.37 \\ 
    Ours - SSIM & -2.13\% & 0.29 & 4.87 & 0.31\% & 0.26 & 5.07 & 2.39\% & 0.18 & 12.06 \\ 
    Ours - PSNR & -2.13\% & 0.22 & 5.53 & 0.31\% & 0.26 & 5.07 & 2.39\% & 0.18 & 12.06 \\ \hline
    \end{tabular}
\vspace{-5mm}
\end{table*}
\subsection{Implementation details}
During the experiment, we consider the most strict setting, where h=1 and t=1 on a ResNet-18 \cite{he2016deep} architecture for three image classification tasks, CIFAR-10 \cite{krizhevsky2009learning}, CIFAR-100 \cite{krizhevsky2009learning}, and a subset of CelebA-HQ \cite{zhu2022celebvhq}. That is, the client only holds the first convolutional layer as well as the last fully connected layer, which is also the minimum requirement for our framework. For CIFAR-10, the intermediate output's feature size is [64x16x16], for CIFAR-100, we remove the MaxPooling layer and the intermediate output's feature size is [64x32x32], and for CelebA, the intermediate output's feature size is [64x64x64]. We consider the ensembled network to contain 10 neural networks (N=10), each being a ResNet-18. The selector secretly selects \{4,3,5\} out of the 10 nets (P=\{4,3,5\}), respectively. The adversarial server is aware of the architecture and the training dataset. It constructs a shadow network $\tilde{\mathbf{M}}_{c,h}$ consisting of three convolutional layers with 64 channels each, with the first one simulating the unknown $\mathbf{M}_{c,h}$, and the other two simulating the Gaussian noise added to the intermediate output. It also has $\tilde{\mathbf{M}}_{c,t}$ with the same shape as $\mathbf{M}_{c,t}$. An adaptive shadow network learns from all 10 server nets with an additional activation layer that is identical to the selector. For any noises added to the intermediate outputs during the training and inference stage, we consider a fixed Gaussian noise $\mathbf{g} \sim N(0,0.1)$.

\subsection{Experiment Setup}
%To evaluate the effectiveness of our approach, 
We employ two key metrics: Structural Similarity (SSIM) and Peak Signal Noise Ratio (PSNR). Higher SSIM and PSNR values indicate better reconstruction quality and thus worse defense. In addition, we report the changes in classification accuracy ($\Delta Acc$) after adding the defense mechanism. As our proposed architecture operates in parallel with existing perturbation methods, we mainly compare our method with its single network counterpart (Single) \cite{dwork2006}, where only one $M^i: \{M_{c,h}, M_{s}^i, M_{c,t}\}$ is used. In the ablation study, we provide a deeper analysis using CIFAR-10, and we compare the following methods against \textit{Ensembler}: no protection (None), training the noise injected to Single (Shredder \cite{shredder}), adding dropout layer in the single network or ensembled network, but without the first stage training (DR-single and DR-10) \cite{dropout}. For the proposed architecture, we evaluate the performance of both reconstruction attacks discussed in Section \ref{intuition}. For reconstruction with a single server net, we report the best reconstruction, or worst defense, in terms of SSIM (Ours - SSIM) and PSNR (Ours - PSNR) among the $N$ networks. For the reconstruction using all networks, we denote it as (Ours - Adaptive). We run the experiments on an A6000 GPU server using Python and PyTorch. For Section 4, we used a mixture of the server with A6000 and Google Colab with T4 GPU. In addition, we report the real-world time delay of the proposed method compared with its single counterpart and a recent encryption-based method \cite{huang2022efficient}. We evaluate the inference time on a Raspberry Pi communicating with the A6000 server over a wired network.

\subsection{Comparison of Results}
%\textcolor{red}{TODO: rewrite this}

We provide the quantitative evaluations of \textit{Ensembler} compared to its non-ensembled counterpart (Single \cite{dwork2006}) in Table \ref{summary_all}. It could be seen that the proposed framework significantly decreases the reconstruction quality of the adversarial party. \textit{Ensembler} incurs a 2.13\% drop in classification accuracy compared to the model without any protection, which is acceptable compared to its advantage in protecting the privacy of the client's raw input. It should be noted that the adaptive MIA strategy performs worse than the best reconstruction with a single network, which is expected. While Equation \ref{train2} constraints $M_{c,h}$ to learn from all server networks, it might still have one that is more ``favored''. Thus, a reconstruction based on that favored network yields the best results. On the other hand, a reconstruction based on all networks might not select the same one as its ``favored'' network, which causes the shadow network's weights to significantly deviate from $M_{c,h}$.

\begin{table}[t]
    \centering
    \tabcolsep 12pt
    \caption{Different defense mechanisms with CIFAR-10. The last three are the proposed framework. The underscored number represents the best defense from other existing methods.}
    \label{summary}
    \begin{tabular}%{p{0.12\textwidth}p{0.08\textwidth}p{0.08\textwidth}p{0.08\textwidth}}
    {lccc}
    \hline
        Name & $\Delta$ Acc & SSIM $\downarrow$ & PSNR $\downarrow$ \\ \hline
        None & 0.00\% & 0.49 & 9.86 \\ 
        Shredder & -2.92\% & \underline{0.29} & 6.70 \\ 
        Single & 2.15\% & 0.39 & 7.53 \\ 
        DR-single & 2.70\% & 0.35 &\underline{ 6.67} \\ 
        DR-10 - SSIM & 1.42\% & 0.37 & 7.35 \\ 
        DR-10 - PSNR & 1.42\% & 0.32 & 7.96 \\ 
        \hline
        Ours - Adaptive & -2.13\% & 0.06 & 5.98 \\ 
        Ours - SSIM & -2.13\% & 0.29 & 4.87 \\ 
        Ours - PSNR & -2.13\% & 0.22 & 5.53 \\ \hline
    \end{tabular}
    \vspace{-3mm}
\end{table}

Moreover, we delve into a more thorough analysis of \textit{Ensembler}'s performance on the CIFAR-10 dataset compared to other existing methods. The results are summarized in Table \ref{summary}. It could be observed that \textit{Ensembler} defends the client's input on par with Shredder \cite{shredder} on structural similarity, and exceeds it on peak signal-noise ratio. In addition, it should be noted that Shredder \cite{shredder} and dropout defense \cite{dropout} can be combined with \textit{Ensembler} together. The additive noise $N(0,\sigma)$ in the third stage could be replaced by Shredder's trained noise, and dropout can also be added to the network's FC layer to perform further protection of the client's sensitive data.

%We provide the quantitative evaluations for CIFAR-10 in Table. \ref{summary}, and the visual assessments in Figure \ref{fig:figure_summ}. It could be seen that the proposed framework significantly increases the reconstruction difficulty of the adversarial party. \textit{Ensembler} incurs a 2.13\% drop in classification accuracy compared to the model without any protection, which is marginal compared to its advantage in protecting privacy of the client's raw input. From the figure, it is clear that the reconstructed images are hardly recognizable by human-level interpretations.

%In addition, we provide the quantitative evaluations for CIFAR-100 and CelebA-HQ in Table. \ref{summary2} and \ref{summary3}, and the visual assessments of the CelebA-HQ dataset in Figure \ref{fig:figure_celeb}. The proposed framework remains effective when the feature size increases. In particular, the framework safeguards the model's prediction ability while protecting the input images on par with the random head network. Although the visual assessments show that increasing feature size leads to better visual recognition, we argue that it is inevitable with simple Gaussian noises. In particular, the shadow network is able to raise the reconstruction quality of a totally mismatched random $M_{c,h}$ to beyond human-recognition level from the shadow network with best PSNR. 

%One more important observation is that the Adaptive attack that uses all N nets perform much worse than using a single network. This is expected, as discussed in 4.4. 

\subsection{Latency}
%2.68
Time required by the standard CI setting and \textit{Ensembler} is provided in Table \ref{time_latency}. In addition, STAMP \cite{huang2022efficient} is a hardware-optimized version of encryption-based private inference, and we show their reported results (LAN-GPU). For all experiments, we use the setting in STAMP to run ResNet-18 on a batch size of 128 images. \textit{Ensembler} incurs only \textbf{0.19 seconds (4.8\%)} overhead compared with running the standard CI inference pipeline, but with much stronger protection on the client's sensitive data. It can also be observed that the increment of time comes from two parts, with more server computing taking the minor role and more communication between the client and the server taking the major role. In the future, more attention on improving client-server communication is pivotal for a more efficient and practical CI inference.

\begin{table}[t]
    \centering
    \tabcolsep 8pt
    \caption{Time Taken (seconds) to run a batch of 128 images through different defense strategies.}
    \label{time_latency}
    \begin{tabular}%{p{0.12\textwidth}p{0.08\textwidth}p{0.08\textwidth}p{0.08\textwidth}}
    {lcccc}
    \hline
        Name & Client & Server & Communication & Total \\ \hline
        Standard CI & 0.66 & 0.98 & 2.30 & 3.94 \\ 
        \textit{Ensembler} & 0.66 & 1.02 & 2.45 & 4.13 \\ 
        STAMP \cite{huang2022efficient} & -- & -- & -- & 309.7\\ 
        \hline
        \end{tabular}
        \vspace{-3mm}

    \end{table}
\section{Conclusion}
\label{conclusion}
During collaborative inference, the client's sensitive information is prone to model inversion attacks from the adversarial server, especially when the client only holds a very small portion of the network. In this paper, we present a novel collaborative inference framework, \textit{Ensembler}, designed to significantly increase the complexity of reconstruction for adversarial parties. \textit{Ensembler} leverages a selective model ensemble on the adversarial server and a secret Selector to defend against the server's model inversion attacks. In addition, it seamlessly aligns with existing methods that introduce diverse forms of noise to intermediate outputs, potentially yielding robust and adaptable architectures if combined with them. Lastly, \textit{Ensembler} is a latency-friendly framework that incurs only marginal overhead during inference. Our experiments highlight the substantial deterioration in reconstruction quality for images safeguarded by \textit{Ensembler} when compared to those without its protection.

\renewcommand{\bibfont}{\small}

\bibfont{
\bibliographystyle{ieeetran}
\bibliography{mybibfile}
}

\end{document}